\documentclass{article} 
\usepackage{amsmath}
\usepackage{amsfonts}
\usepackage{amssymb}
\usepackage[dvipsone]{graphicx}
\usepackage{subfig}

\title{Computation of Object Approach by a Biophysical Model of a Wide-Field Visual Neuron: Dynamics, Peaks, \& Fits}

\author{
Matthias S. Keil\thanks{\texttt{http://www.ir3c.ub.edu}, Research Institute for Brain, Cognition, and Behaviour (IR3C)
Edifici de Ponent, Campus Mundet, Universitat de Barcelona, Passeig Vall d'Hebron, 171. E-08035 Barcelona.
Note: A revised version of this paper with the title ``\emph{Emergence of Multiplication in a Biophysical
Model of a Wide-Field Visual Neuron for Computing Object Approaches: Dynamics, Peaks, \& Fits}'' has been
accepted in \emph{Advances in Neural Information Processing Systems} \textbf{NIPS 2011}, Granda, Spain
(\texttt{http://nips.cc}).}\\
Department of Basic Psychology\\
University of Barcelona\\
E-08035 Barcelona, Spain\\
}

\begin{document}

\maketitle

\begin{abstract}
Many species show avoidance reactions in response to looming object approaches.
In locusts, the corresponding escape behavior correlates with the activity
of the lobula giant movement detector (LGMD) neuron.  During an object approach,
its firing rate was reported to gradually increase until a peak is reached,
and then it declines quickly.  The $\eta$-function predicts that the LGMD activity
is a product between an exponential function of angular size $\exp(-\Theta)$ and
angular velocity $\dot{\Theta}$, and that peak activity is reached before time-to-contact
(ttc).  The $\eta$-function has become the prevailing LGMD model because it
reproduces many experimental observations, and even experimental evidence for
the multiplicative operation was reported.  Several inconsistencies remain
unresolved, though.  Here we address these issues with a new model ($\psi$-model),
which explicitly connects $\Theta$ and $\dot{\Theta}$ to biophysical quantities.
The $\psi$-model avoids biophysical problems associated with implementing
$\exp(\cdot)$, implements the multiplicative operation of $\eta$ via divisive
inhibition, and explains why activity peaks could occur after ttc.  It consistently
predicts response features of the LGMD, and provides excellent fits to published
experimental data, with goodness of fit measures comparable to corresponding
fits with the $\eta$-function.
\end{abstract}
%
\section{Introduction: $\tau$ and $\eta$}
%
Collision sensitive neurons were reported in species such different as
monkeys \cite{FogassiEtAl1996,CookeGraziano2004},
pigeons \cite{WangFrost92,SunFrost98},
frogs \cite{KingLettvinGruberg1999,NakagawaHongjian10}, and insects
\cite{Schlotterer1977, RindSimmons92a,RindSimmons92b,HatsopoulosEtAl95,WickleinStrausfeld2000}.
This indicates a high ecological relevance, and raises the question about how neurons compute
a signal that eventually triggers corresponding movement patterns (e.g. escape behavior or
interceptive actions).  Here, we will focus on visual stimulation.  Consider, for
simplicity, a circular object (diameter $2l$), which approaches the eye at a collision course
with constant velocity $v$.  If we do not have any \emph{a priori} knowledge about the object
in question (e.g. its typical size or speed), then we will be able to access only two information
sources.  These information sources can be measured at the retina and are called optical variables
(OVs).  The first is the visual angle $\Theta$, which can be derived from the number of stimulated
photoreceptors (spatial contrast).  The second is its rate of change $d\Theta(t)/dt \equiv\dot{\Theta}(t)$.
Angular velocity $\dot{\Theta}$ is related to temporal contrast.\\
How should we combine $\Theta$ and $\dot{\Theta}$ in order to track an imminent collision?
The perhaps simplest combination is $\tau(t)\equiv\Theta(t)/\dot{\Theta}(t)$ \cite{Hoyle1957,Lee1976}.
If the object hit us at time $t_c$, then $\tau(t)\approx t_c-t$ will give us a
running estimation of the time that is left until contact\footnote{This linear approximation
gets worse with increasing $\Theta$, but turns out to work well until short before ttc.}.
Moreover, we do not need to know anything about the approaching object: The ttc estimation computed
by $\tau$ is independent of object size and velocity.  Neurons with $\tau$-like
responses were indeed identified in the nucleus retundus of the pigeon brain \cite{SunFrost98}.
In humans, only fast interceptive actions seem to rely exclusively on $\tau$ \cite{Wann1976,Tresilian99}.
Accurate ttc estimation, however, seems to involve further mechanisms (rate of disparity change \cite{RushtonWann1999}).\\
Another function of OVs with biological relevance is $\eta\equiv\dot{\Theta}\exp(-\alpha\Theta)$,
with $\alpha=const.$ \cite{HatsopoulosEtAl95}.  While $\eta$-type neurons were found again in
pigeons \cite{SunFrost98} and bullfrogs \cite{NakagawaHongjian10}, most data were gathered
from the LGMD\footnote{LGMD activity is usually monitored via its postsynaptic neuron, the Descending Contralateral
Movement Detector (DCMD) neuron.  This represents no problem as LGMD spikes follow DCMD spikes
1:1 under visual stimulation \cite{OSheaWilliams74} from 300Hz \cite{OSheaRowell75} to at least 400Hz \cite{Rind84}.}
in locusts (e.g. \cite{HatsopoulosEtAl95,GabbianiKrappLaurent99,GabbianiEtAl2004,PeronGabbiani09}).
The $\eta$-function is a phenomenological model for the LGMD, and implies three principal
hypothesis:
\textit{(i)} An implementation of an exponential function $\exp(\cdot)$.  Exponentation is thought
to take place in the LGMD axon, via active membrane conductances \cite{GabKraKocLau02}.
Experimental data, though, seem to favor a third-power law rather than $\exp(\cdot)$.
\textit{(ii)} The LGMD carries out biophysical computations for implementing the multiplicative operation.
It has been suggested that multiplication is done within the LGMD itself, by subtracting the logarithmically
encoded variables $\log\dot{\Theta}-\alpha\Theta$ \cite{HatsopoulosEtAl95,GabKraKocLau02}.
\textit{(iii)} The peak of the $\eta$-function occurs before ttc, at visual angle $\Theta(\hat{t})=2\arctan(1/\alpha)$
\cite{GabbianiKrappLaurent99}.  It follows ttc for certain stimulus configurations (e.g. $l/|v|\lessapprox 5ms$).
In principle, $\hat{t}>t_c$ can be accounted for by $\eta(t+\delta)$ with a fixed delay $\delta<0$ (e.g. $-27ms$).
But other researchers observed that LGMD activity continuous to rise after ttc even for $l/|v|\gtrapprox 5ms$ \cite{RindSimmons97}.
These discrepancies remain unexplained so far \cite{RindSimmons99b}, but stimulation dynamics perhaps plays a role.\\
We we will address these three issues by comparing the novel function ``$\psi$'' with the $\eta$-function.
%
\section{LGMD computations with the $\psi$-function}
%
A circular object which starts its approach at distance $x_0$ and with speed $v$ projects a visual
angle $\Theta(t)=2\arctan[l/(x_0-vt)]$ on the retina \cite{SunFrost98,GabbianiKrappLaurent99}.
The kinematics is hence entirely specified by the half-size-to-velocity ratio $l/|v|$, and $x_0$.
Furthermore, $\dot{\Theta}(t)=2lv/((x_0-vt)^2+l^2)$.\\
In order to define $\psi$, we consider at first the LGMD neuron as an RC-circuit
with membrane potential\footnote{Here we assume that the membrane potential serves as a predictor
for the LGMD's mean firing rate.} $V$ \cite{Kochbuch99}
\def\gleak{\beta}
\begin{equation}\label{membrane}
  C_m\frac{dV}{dt}=\gleak\left(V_{rest}-V\right) + g_{exc}\left(V_{exc}-V\right) + g_{inh}\left(V_{inh}-V\right)
\end{equation}
$C_m$ $=$ membrane capacity\footnote{Set to unity for all simulations}; $\gleak\equiv 1/R_m$ denotes leakage conductance
across the cell membrane ($R_m$: membrane resistance); $g_{exc}$ and $g_{inh}$ are excitatory and inhibitory inputs.
Each conductance $g_i$ ({\small $i=\mathit{rest}, \mathit{exc}, \mathit{inh}$}) can drive the membrane
potential to its associated reversal potential $V_i$ (usually $V_{inh}\leq V_{exc}$).
Shunting inhibition means $V_i=V_{rest}$.  Shunting inhibition lurks ``silently'' because it gets effective only if the
neuron is driven away from its resting potential.  With synaptic input, the neuron decays into its equilibrium state
$V_{\infty}\equiv(V_{rest}\gleak+V_{exc}g_{exc}+V_{inh}g_{inh})/(\gleak+g_{exc}+g_{inh})$ according to
$V(t)=V_{\infty}(1-exp(-t/\tau_m))$.  Without external input, $V(t\gg1) \rightarrow V_{rest}$.
The time scale is set by $\tau_m$.  Without synaptic input $\tau_m\equiv C_m/\gleak$.  Slowly varying inputs
$g_{exc},g_{inh}>0$ modify the time scale to approximately $\tau_m/(1+(g_{exc}+g_{inh})/\gleak)$.  For highly
dynamic inputs, such as in late phase of the object approach, the time scale gets dynamical as well.
The $\psi$-model assigns synaptic inputs\footnote{LGMD receives also inhibition from a laterally acting
network \cite{OSheaRowell75}.  The $\eta$-function considers only direct feedforward inhibition
\cite{OSheaWilliams74,GabbianiCohenLaurent2005}, and so do we.}
\begin{subequations}\label{psi}
\begin{align}
g_{exc}(t) &= \dot{\vartheta}(t), 		&  \dot{\vartheta}(t)=\zeta_1\dot{\vartheta}(t-\Delta t_{stim}) + (1-\zeta_1)\dot{\Theta}(t)	\label{exc}\\
g_{inh}(t) &= \left[\gamma\vartheta(t)\right]^e, &  \vartheta(t)=\zeta_0\vartheta(t-\Delta t_{stim}) + (1-\zeta_0)\Theta(t) 			\label{inh}
\end{align}
\end{subequations}
Thus, we say $\psi(t)\equiv V(t)$ if and only if $g_{exc}$ and $g_{inh}$ are defined with the last equation.
The time scale of stimulation is defined by $\Delta t_{sim}$ (by default $1ms$).
The variables $\vartheta$ and $\dot{\vartheta}$ are lowpass filtered angular size and rate of expansion, respectively.
The amount of filtering is defined by memory constants $\zeta_0$ and $\zeta_1$ (no filtering for zero).  The idea is
to continue to generate synaptic input after ttc, where $\Theta(t>t_c)=const$ and thus $\dot{\Theta}(t>t_c)=0$.
Inhibition is first weighted by $\gamma$, and then potentiated by the exponent $e$.  Hodgkin-Huxley potentiates
gating variables $n, m \in [0,1]$ instead (potassium $\propto n^4$, sodium $\propto m^3$, \cite{HodkinHuxley1952}) and
multiplies them with conductances.
Gabbiani and co-workers found that the function which transforms membrane potential to firing rate is better described by
a power function with $e=3$ than by $\exp(\cdot)$ (Figure 4d in \cite{GabKraKocLau02}).
\begin{figure}[t!]
 \centering
  \subfloat[discretized optical variables]{\includegraphics[width=0.475\linewidth]{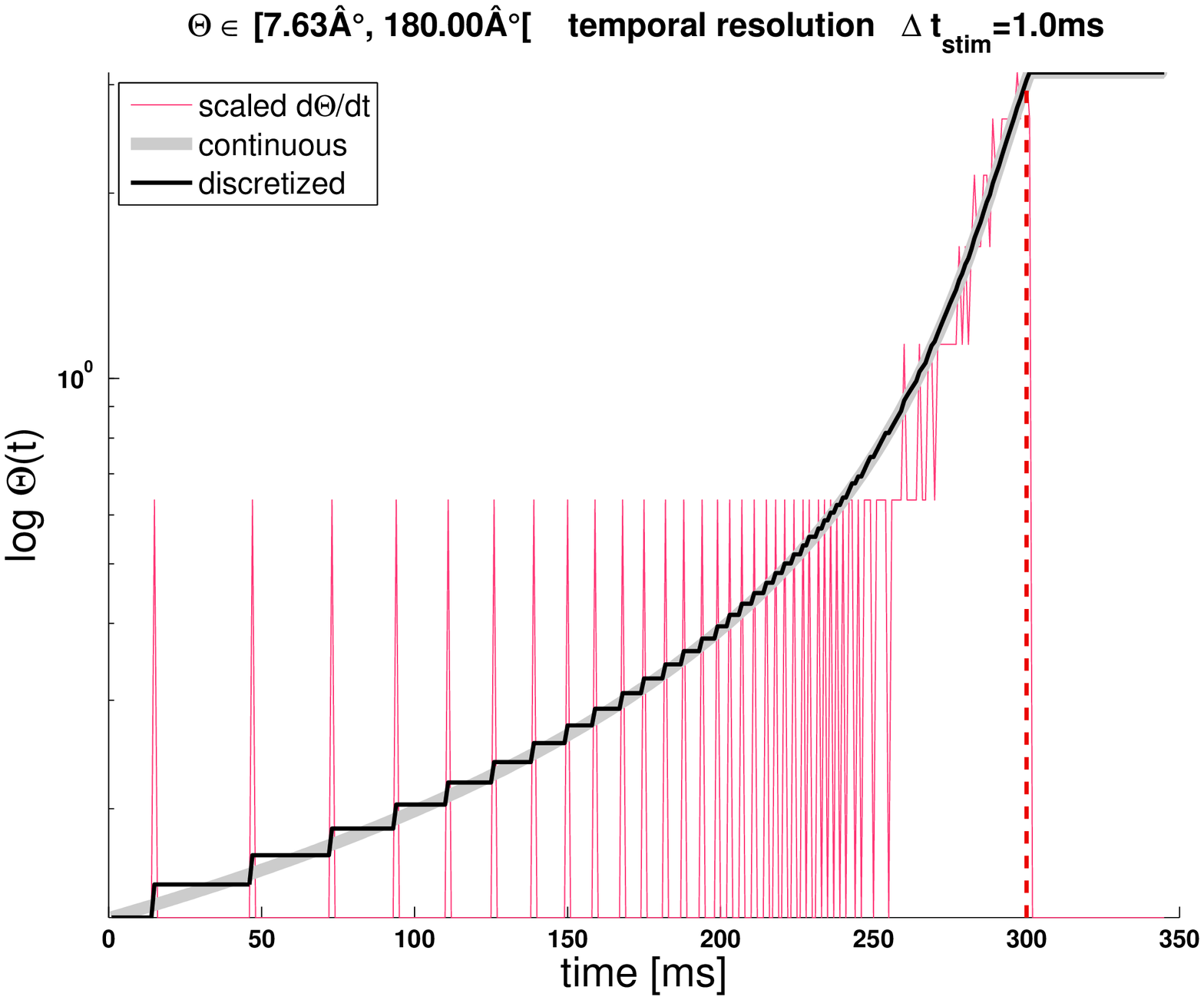}}~
  \subfloat[$\psi$ versus $\eta$]{\includegraphics[width=0.475\linewidth]{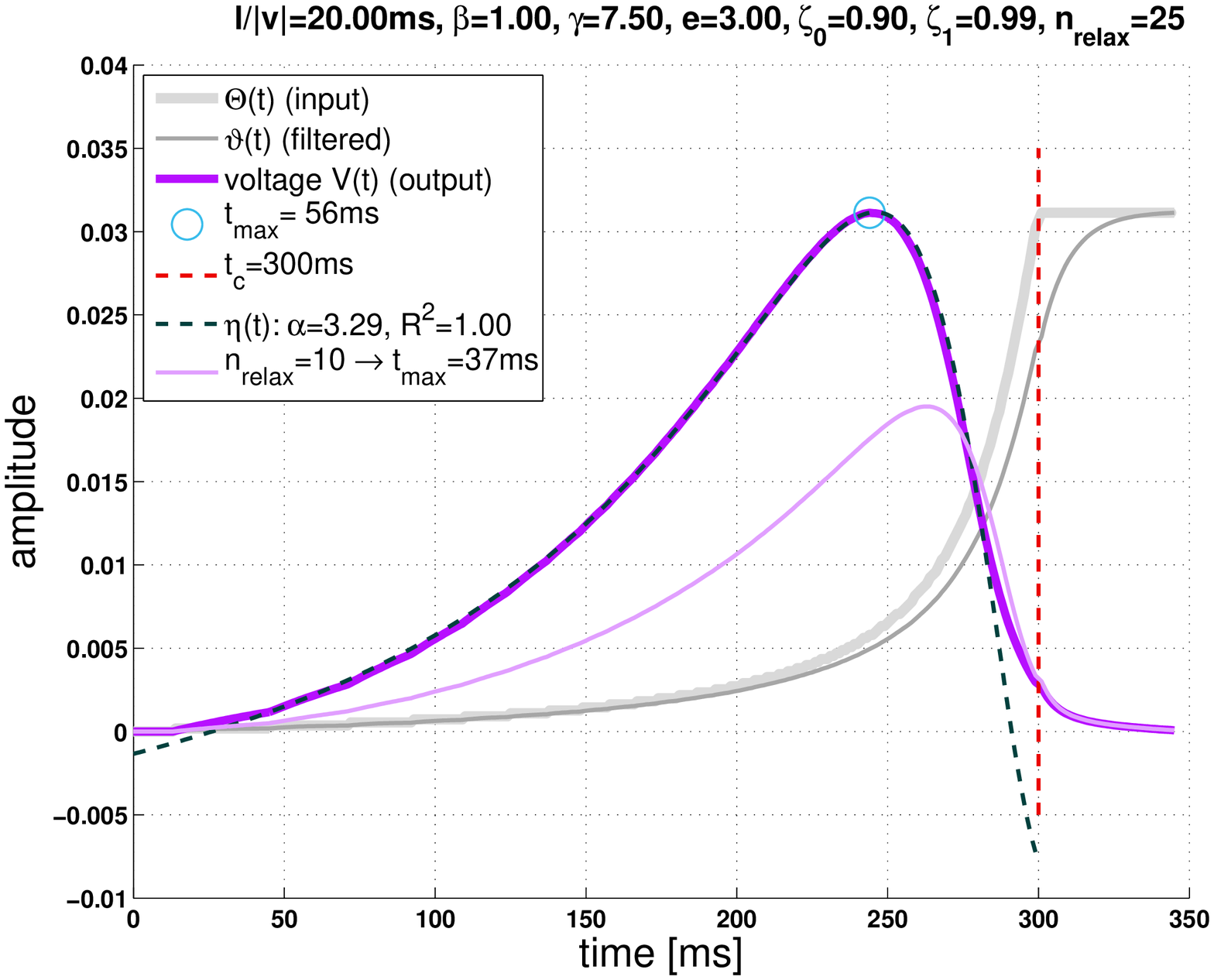}}
\caption{\label{VoltageTrace}\textit{(a)} The continuous visual angle of an approaching object is shown
along with its discretized version.  Discretization transforms angular velocity from a continuous variable
into a series of ``spikes'' (rescaled).
\textit{(b)} The $\psi$ function with the inputs shown in \textit{a}, with $n_{relax}=25$ relaxation time
steps.  Its peak occurs $t_{max}=56ms$ before ttc ($t_c=300ms$).  An $\eta$ function ($\alpha=3.29$) that
was fitted to $\psi$ shows good agreement.  For continuous optical variables, the peak would occur $4ms$
earlier, and $\eta$ would have $\alpha=4.44$ with $R^2=1$.  For $n_{relax}=10$, $\psi$ is farther away from
its equilibrium at $V_{\infty}$, and its peak moves $19ms$ closer to ttc.}
\end{figure}
\begin{figure}[t!]
 \centering
  \subfloat[different $n_{relax}$]{\includegraphics[width=0.475\linewidth]{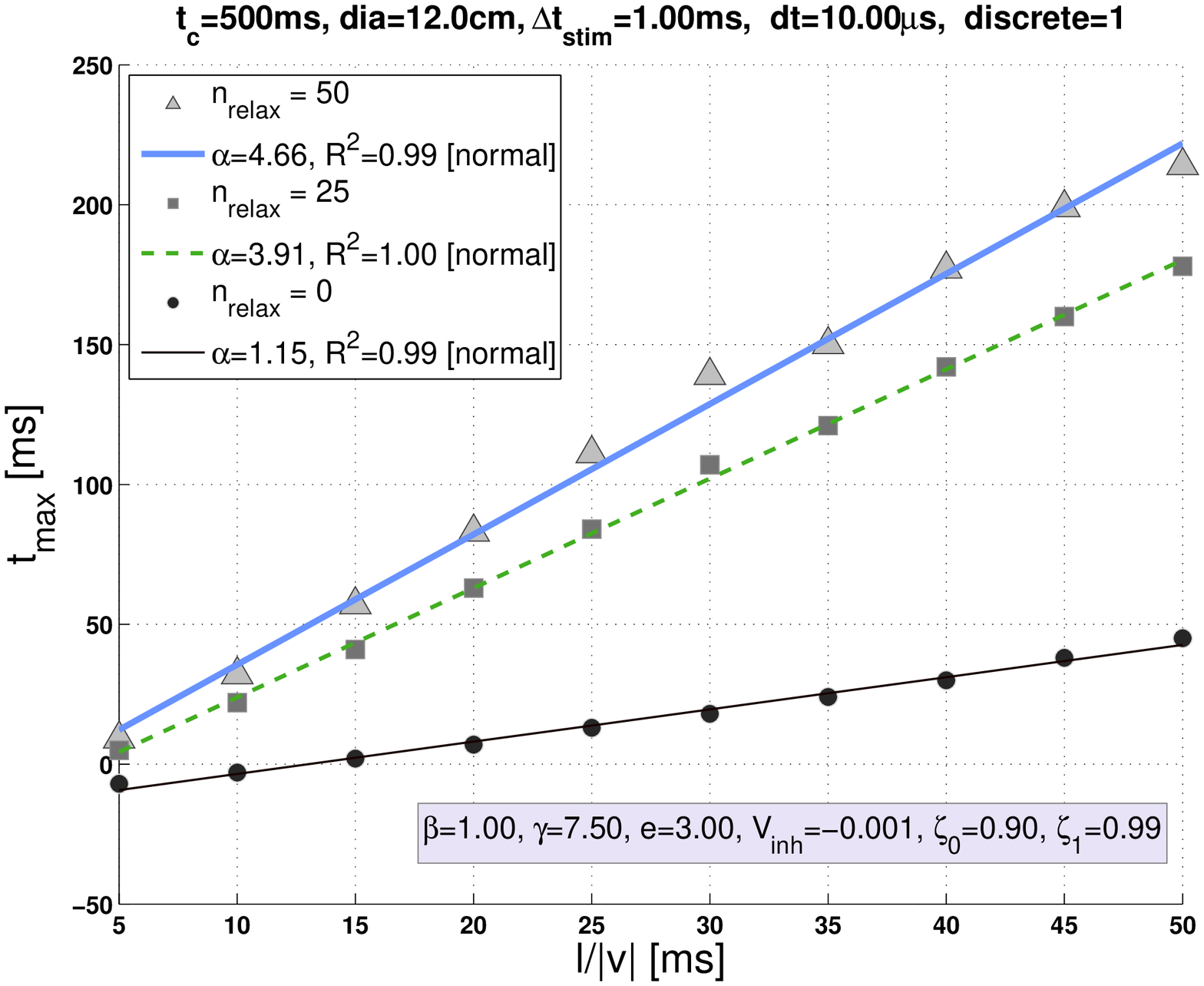}}~
  \subfloat[different $\Delta t_{stim}$]{\includegraphics[width=0.475\linewidth]{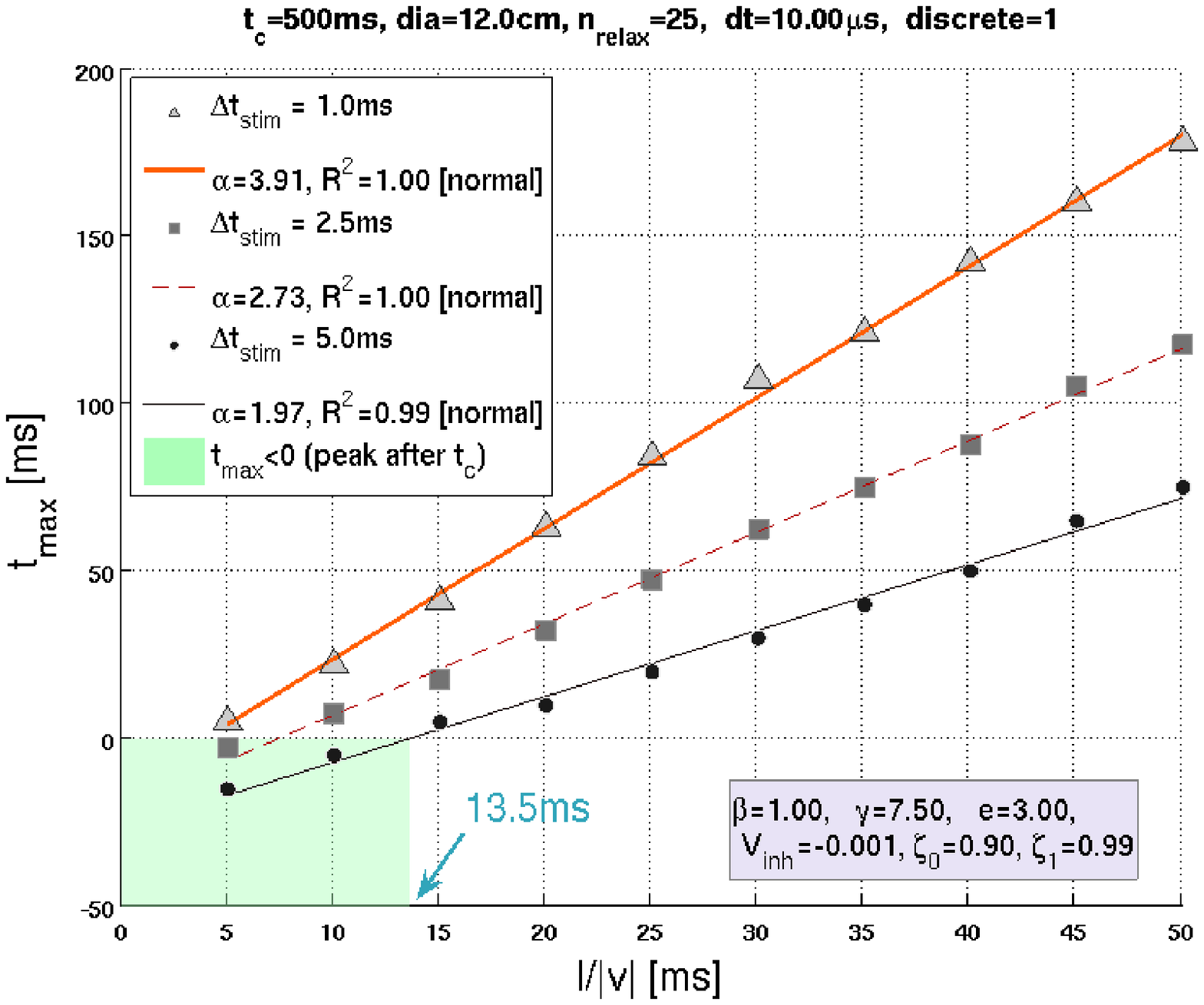}}
\caption{\label{integratePSI}
The figures plot the relative time $t_{max}\equiv t_c-\hat{t}$ of the response peak of $\psi$, $V(\hat{t})$,
as a function of half-size-to-velocity ratio (points).  Line fits with slope $\alpha$ and intercept $\delta$ were
added (lines).  The predicted linear relationship in all cases is consistent with experimental evidence
\cite{GabbianiKrappLaurent99}.
\textit{(a)} The stimulus time scale is held constant at $\Delta t_{stim}=1ms$, and several LGMD time scales are defined
by $n_{relax}$.  Bigger values of $n_{relax}$ move $V(t)$ closer to its equilibrium $V_{\infty}(t)$, implying
higher slopes $\alpha$ in turn.
\textit{(b)} LGMD time scale is fixed at $n_{relax}=25$, and  $\Delta t_{stim}$ is manipulated.  Because of the
discretization of optical variables (OVs) in our simulation, increasing $\Delta t_{stim}$ translates to an overall
smaller number of jumps in OVs, but each with higher amplitude.}
\end{figure}
%
\section{Dynamics of the $\psi$-function}
%
%
\textbf{Discretization.} In a typical experiment, a monitor is placed a short distance away form the insect's eye, and
an approaching object is displayed.  Computer screens have a fixed spatial resolution, and as a consequence size increments
of the displayed object proceed in discrete jumps.  The locust retina is furthermore composed of a discrete array of ommatidia
units.  We therefore can expect a corresponding step-wise increment of $\Theta$ with time, although optical
and neuronal filtering may smooth $\Theta$ to some extent again, resulting in $\vartheta$ (figure \ref{VoltageTrace}).
Discretization renders $\dot{\Theta}$ discontinuous, what again will be alleviated in $\dot{\vartheta}$.  For simulating
the dynamics of $\psi$, we discretized angular size with $\operatorname*{floor}(\Theta)$, and $\dot{\Theta}(t)\approx
[\Theta(t-\Delta t_{stim})-\Theta(t)]/\Delta t_{sim}$.
Discretized optical variables (OVs) were re-normalized to match the range of original (i.e. continuous) OVs.\\
\textbf{To peak, or not to peak?} Rind \& Simmons reject the hypothesis that the activity peak signals impending
collision on grounds of two arguments \cite{RindSimmons97}:
\textit{(i)} If $\Theta(t+\Delta t_{stim})-\Theta(t) \gtrapprox 3^o$ in consecutively displayed stimulus frames,
the illusion of an object approach would be lost.  Such stimulation would rather be perceived as a sequence of rapidly
appearing (but static) objects, causing reduced responses.
\textit{(ii)} After the last stimulation frame has been displayed (that is $\Theta=const$), LGMD responses keep on
building up following ttc.  This behavior clearly depends on $l/|v|$, also according to their own data (e.g. Figure 4
in \cite{RindSimmons92a}):
Response build up beyond ttc is typically observed for sufficiently small values of $l/|v|$.  Input into $\psi$ in
situations where $\Theta=const$ and $\dot{Theta}=0$, respectively, is accommodated by $\vartheta$ and $\dot{\vartheta}$,
respectively.\\
We simulated (i) by setting $\Delta t_{stim}=5ms$, thus producing larger and more infrequent jumps
in discrete OVs than with $\Delta t_{stim}=1ms$ (default).  As a consequence, $\vartheta(t)$ grows
more slowly (delayed build up of inhibition), and the peak occurs later ($t_{max}\equiv t_c-\hat{t}=10ms$
with everything else identical with figure \ref{VoltageTrace}b).  The peak amplitude $\hat{V}=V(\hat{t})$
decreases nearly sixfold with respect to default.  Our model thus predicts the reduced responses observed by
Rind \& Simmons \cite{RindSimmons97}.\\
\textbf{Linearity.} Time of peak firing rate is linearly related to $l/|v|$
\cite{HatsopoulosEtAl95,GabbianiKrappLaurent99}.
The $\eta$-function is consistent with this experimental evidence: $\hat{t}=t_c-\alpha l/|v| + \delta$
(e.g. $\alpha=4.7$, $\delta=-27ms$).  The $\psi$-function reproduces this relationship as well
(figure \ref{integratePSI}), where $\alpha$ depends critically on the time scale of biophysical
processes in the LGMD.  We studied the impact of this time scale by choosing $10\mu s$
for the numerical integration of equation \ref{membrane} (algorithm: 4th order Runge-Kutta).  Apart
from improving the numerical stability of the integration algorithm, $\psi$ is far from its
equilibrium $V_{\infty}(t)$ in every moment $t$, given the stimulation time scale $\Delta t_{stim}=1ms$
\footnote{Assuming one $\Delta t_{stim}$ for each integration time step}.  Now, at each
value of $\Theta(t)$ and $\dot{\Theta}(t)$, respectively, we intercalated
$n_{relax}$ iterations for integrating $\psi$.  Each iteration takes $V(t)$ asymptotically closer to
$V_{\infty}(t)$, and $\lim_{n_{relax} \gg 1} V(t)=V_{\infty}(t)$.  If the internal processes in
the LGMD cannot keep up with stimulation ($n_{relax}=0$), we obtain slopes values that underestimate
experimentally found values (figure \ref{integratePSI}a).  In contrast, for $n_{relax}\gtrapprox 25$
we get an excellent agreement with the experimentally determined $\alpha$.  This means that -- under the
reported experimental stimulation conditions (e.g. \cite{GabbianiKrappLaurent99}) -- the LGMD would operate
relatively close to its steady state\footnote{Notice that in this moment we can only make relative
statements - we do not have data at hand for defining absolute time scales}.\\
Now we fix $n_{relax}$ at $25$ and manipulate $\Delta t_{stim}$ instead (figure \ref{integratePSI}b).
The default value $\Delta t_{stim}=1ms$ corresponds to $\alpha=3.91$.  Slightly bigger values
of $\Delta t_{stim}$ ($2.5ms$ and $5ms$) underestimate the experimental $\alpha$.
In addition, the line fits also return smaller intercept values then.  We see  $t_{max}<0$
up to $l/|v|\approx 13.5 ms$ -- LGMD activity peaks after ttc.  Or, in other words,
LGMD activity continues to increase after ttc.  In the limit, where stimulus dynamics is extremely
fast, and LGMD processes are kept far from equilibrium at each instant of the approach, $\alpha$
gets very small.  As a consequence, $t_{max}$ gets largely independent of $l/|v|$: The activity
peak would cling to $t_{max}$ although we varied $l/|v|$.
\begin{figure}[t!]
 \centering
  \subfloat[$\beta$ varies]{\includegraphics[width=0.34\linewidth]{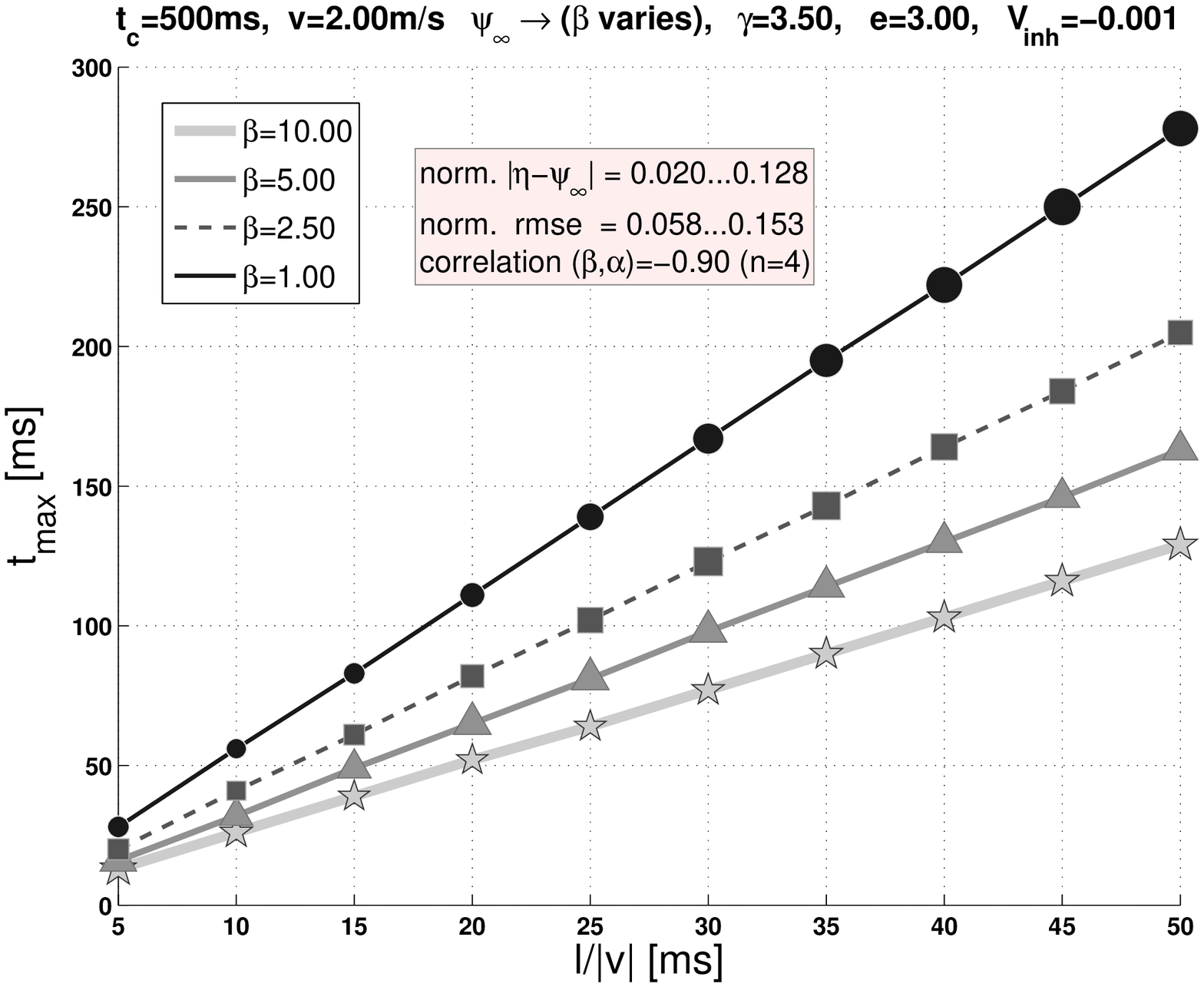}}
  \subfloat[$e$ varies]{\includegraphics[width=0.34\linewidth]{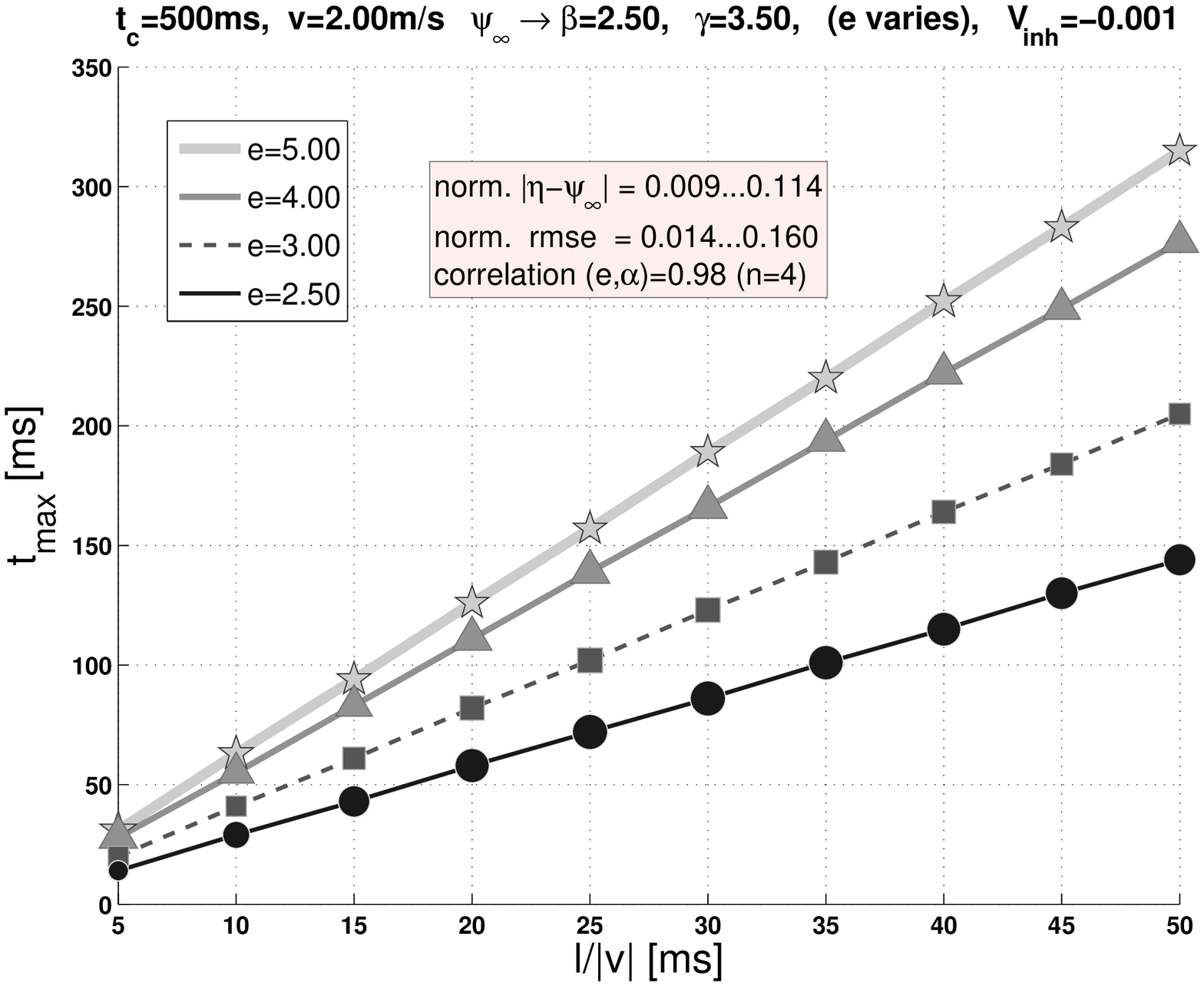}}
  \subfloat[$\gamma$ varies]{\includegraphics[width=0.34\linewidth]{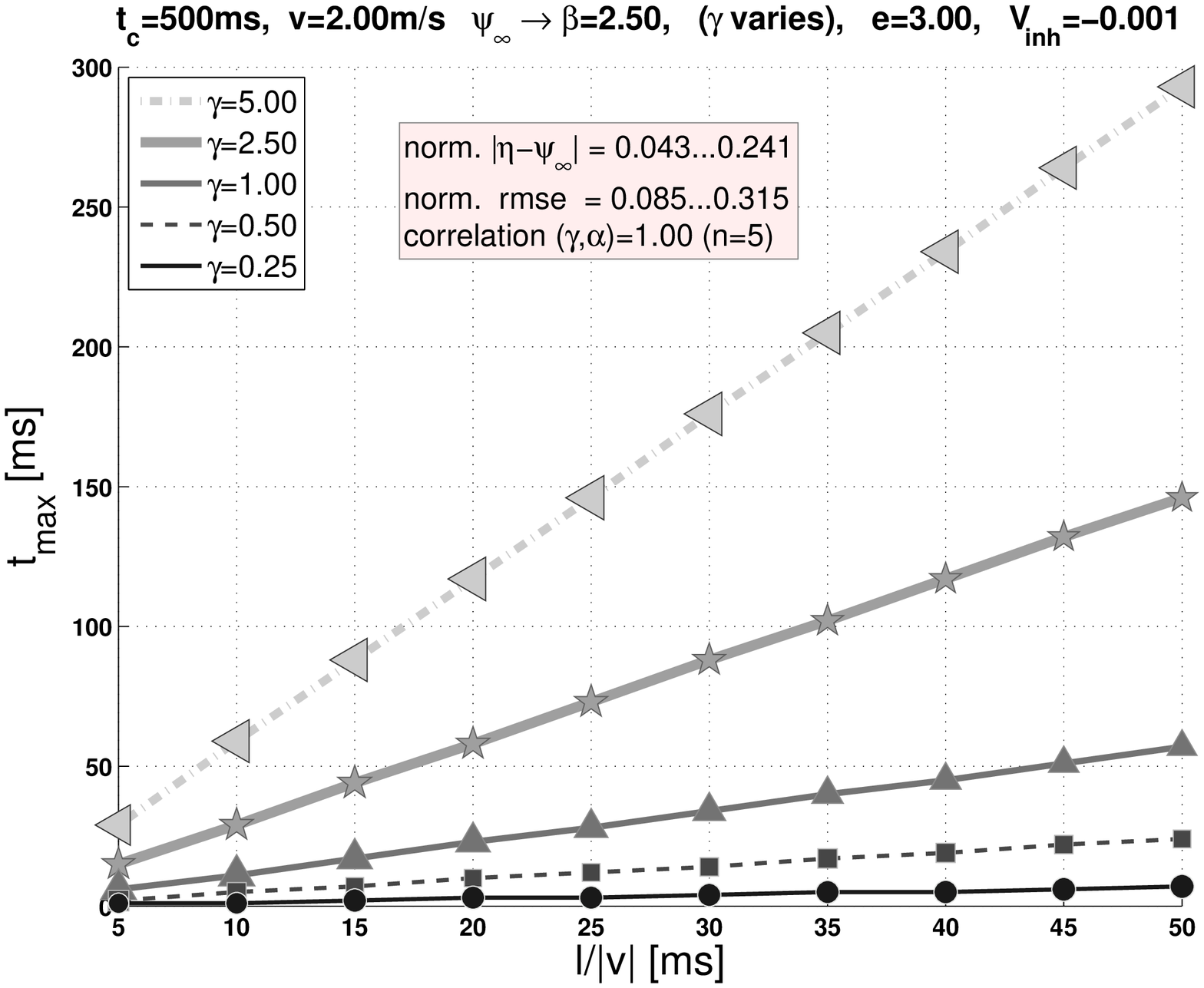}}
\caption{\label{PsiPars}Each curve shows how the peak $\hat{\psi}_{\infty}\equiv\psi_{\infty}(\hat{t})$
depends on the half-size-to-velocity ratio.  In each display, one parameter of $\psi_{\infty}$ is
varied (legend), while the others are held constant (figure title).  Line slopes vary according
to parameter values.  Symbol sizes are scaled according to rmse (see also figure \ref{DiffPsiEta}).
Rmse was calculated between normalized $\psi_{\infty}(t)$ \& normalized $\eta(t)$ (i.e. both
functions $\in[0,1]$ with original minimum and maximum indicated by the textbox).
To this end, the peak of the $\eta$-function was placed at $t_c$, by choosing, at each
parameter value, {\small $\alpha=|v|\cdot (t_c-\hat{t})/l$} (for determining
correlation, the mean value of $\alpha$ was taken across $l/|v|$).}
\end{figure}
\begin{figure}[t!]
 \centering
  \subfloat[$\beta$ varies]{\includegraphics[width=0.34\linewidth]{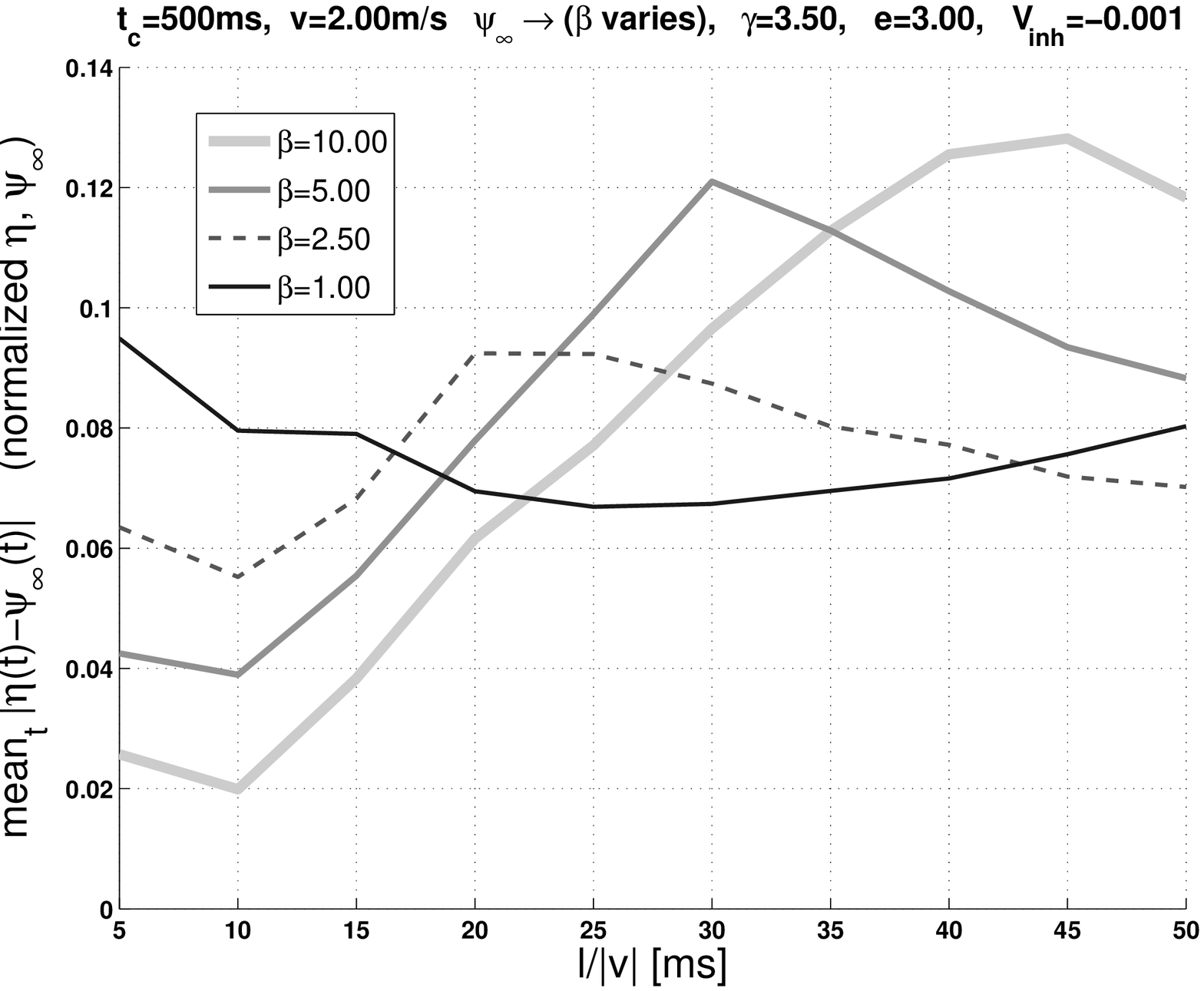}}
  \subfloat[$e$ varies]{\includegraphics[width=0.34\linewidth]{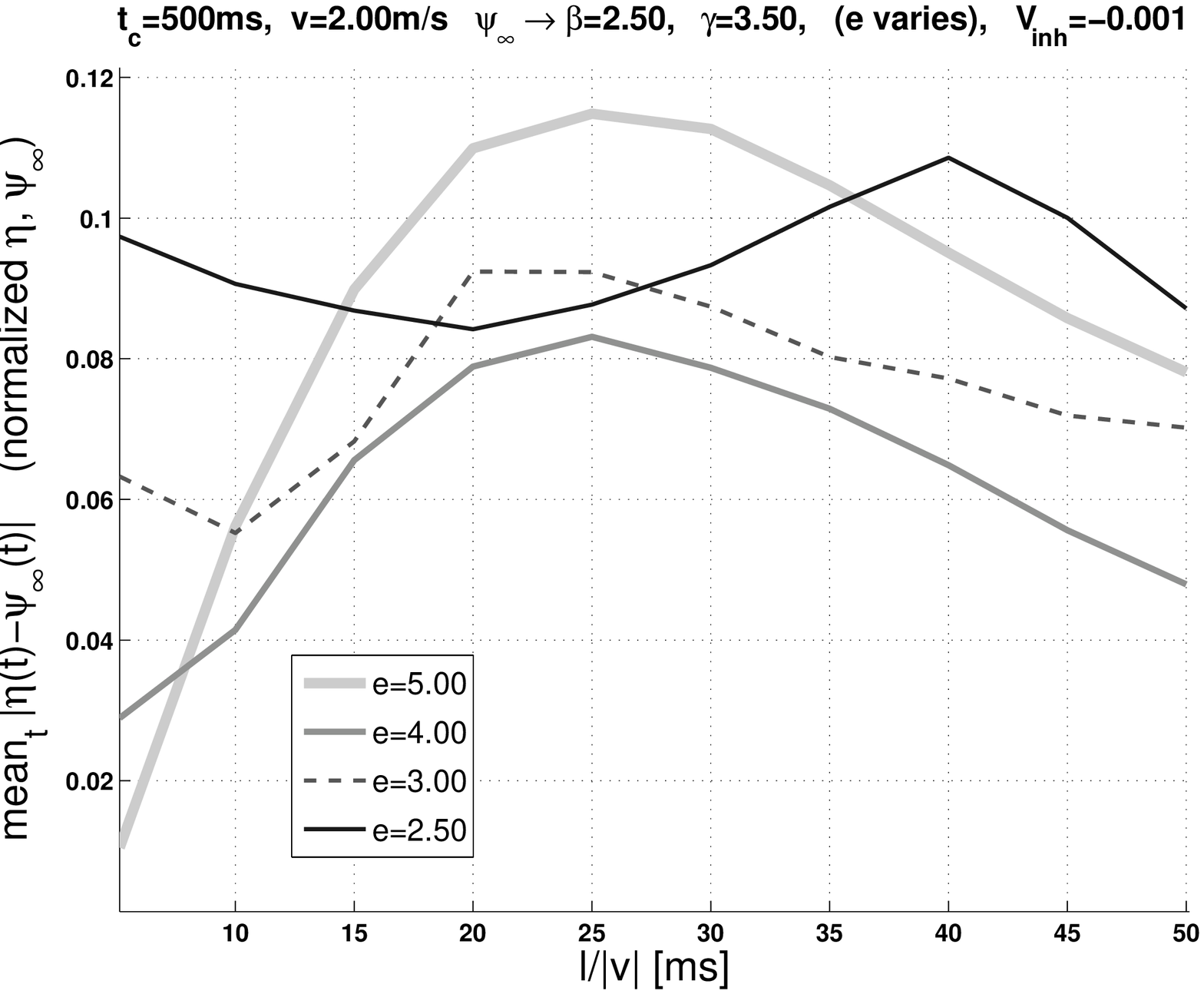}}
  \subfloat[$\gamma$ varies]{\includegraphics[width=0.34\linewidth]{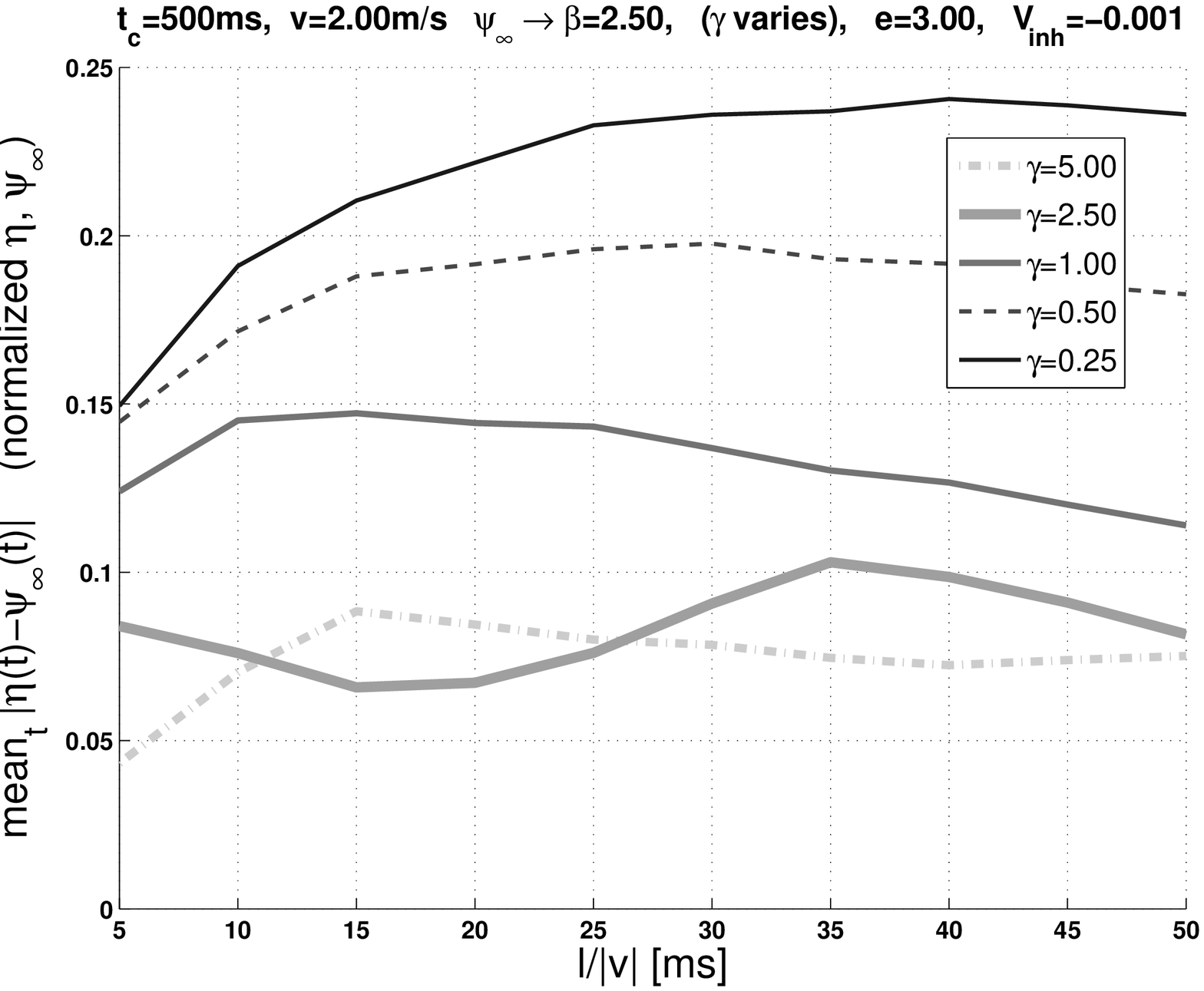}}
\caption{\label{DiffPsiEta}This figure complements figure \ref{PsiPars}.  It visualizes
the time averaged absolute difference between normalized $\psi_{\infty}(t)$ \& normalized $\eta(t)$.
For $\eta$, its value of $\alpha$ was chosen such that the maxima of both functions coincide.
Although not being a fit, it gives a rough estimate on how the shape of both curves deviate from
each other.  The maximum possible difference would be one.
}
\end{figure}
\begin{figure}[t!]
 \centering
  \subfloat[$\dot{\Theta}=126^o/s$]{\includegraphics[width=0.475\linewidth]{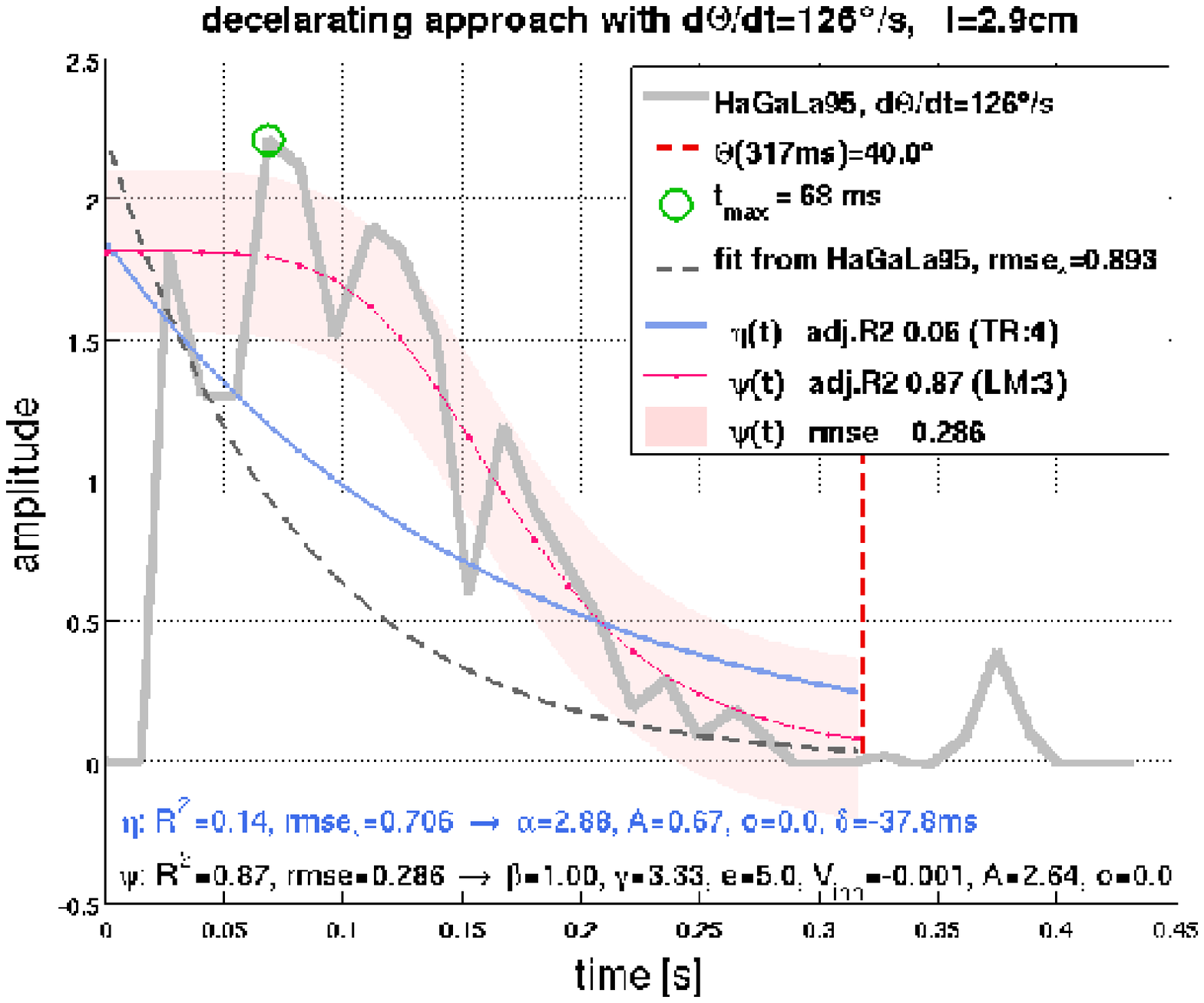}}~
  \subfloat[$\dot{\Theta}= 63^o/s$]{\includegraphics[width=0.475\linewidth]{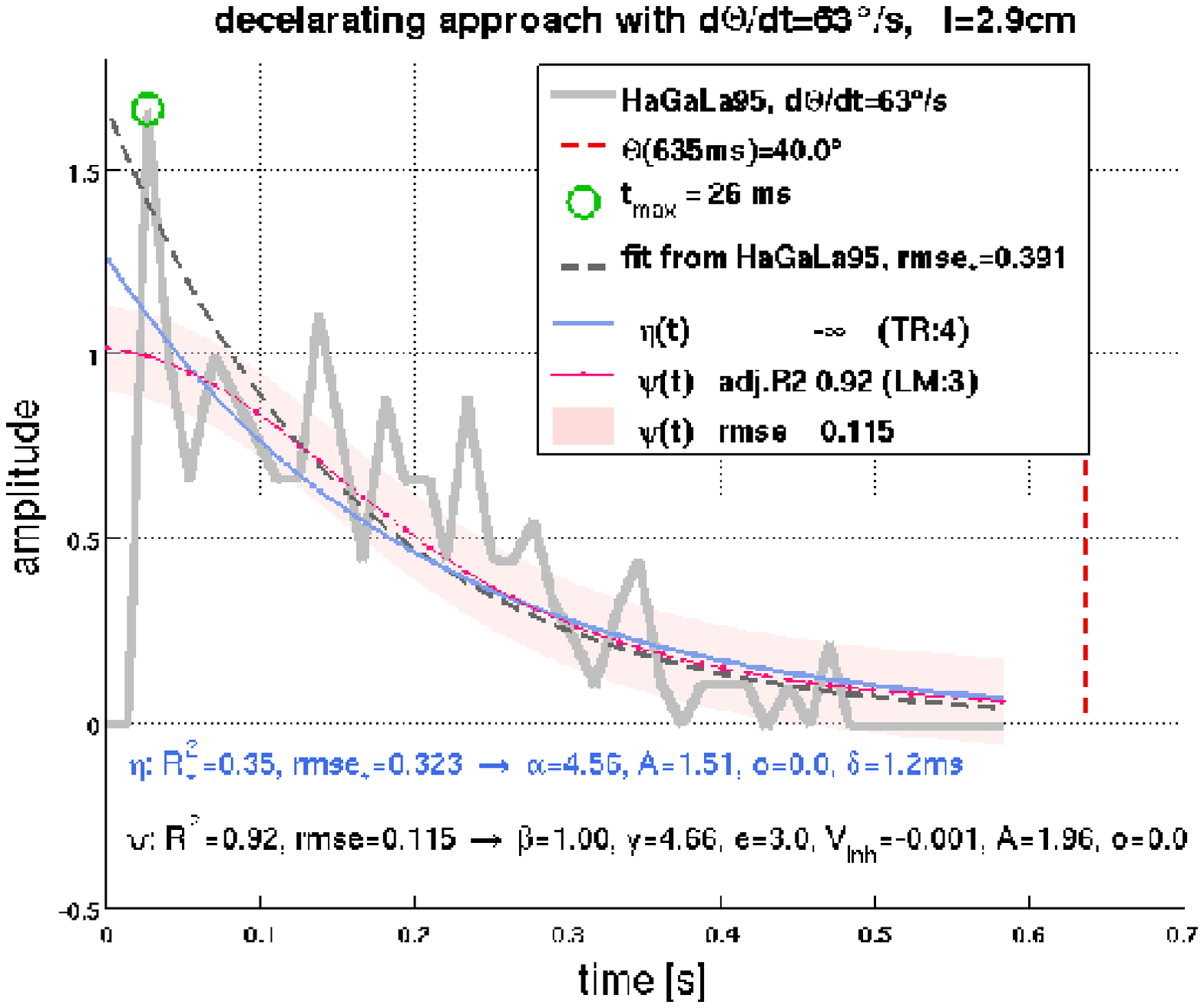}}
\caption{\label{ConstAngularVelocity} The original data (legend label ``HaGaLa95'') were resampled
from ref. \cite{HatsopoulosEtAl95} and show DCMD responses to an object approach with
$\dot{\Theta}=const$.  Thus, $\Theta$ increases linearly with time.  The
$\eta$-function ({\small fitting function: $A\eta(t+\delta)+o$}) and $\psi_{\infty}$
({\small fitting function: $A\psi_{\infty}(t)+o$}) were fitted to these data:
\textit{(a)} (Figure 3 Di in \cite{HatsopoulosEtAl95}) Good fits for $\psi_{\infty}$
are obtained with $e=5$ or higher ({\small $e=3 \rightsquigarrow R^2=0.35$ and $rmse=0.644$;
$e=4 \rightsquigarrow R^2=0.45$ and $rmse=0.592$}).  ``Psi'' adopts a sigmoid-like curve form
which (subjectively) appears to fit the original data better than $\eta$.
\textit{(b)} (Figure 3 Dii in \cite{HatsopoulosEtAl95}) ``Psi'' yields an excellent
fit for $e=3$.}
\end{figure}
\begin{figure}[t!]
 \centering
  \subfloat[spike trace]{\includegraphics[width=0.475\linewidth]{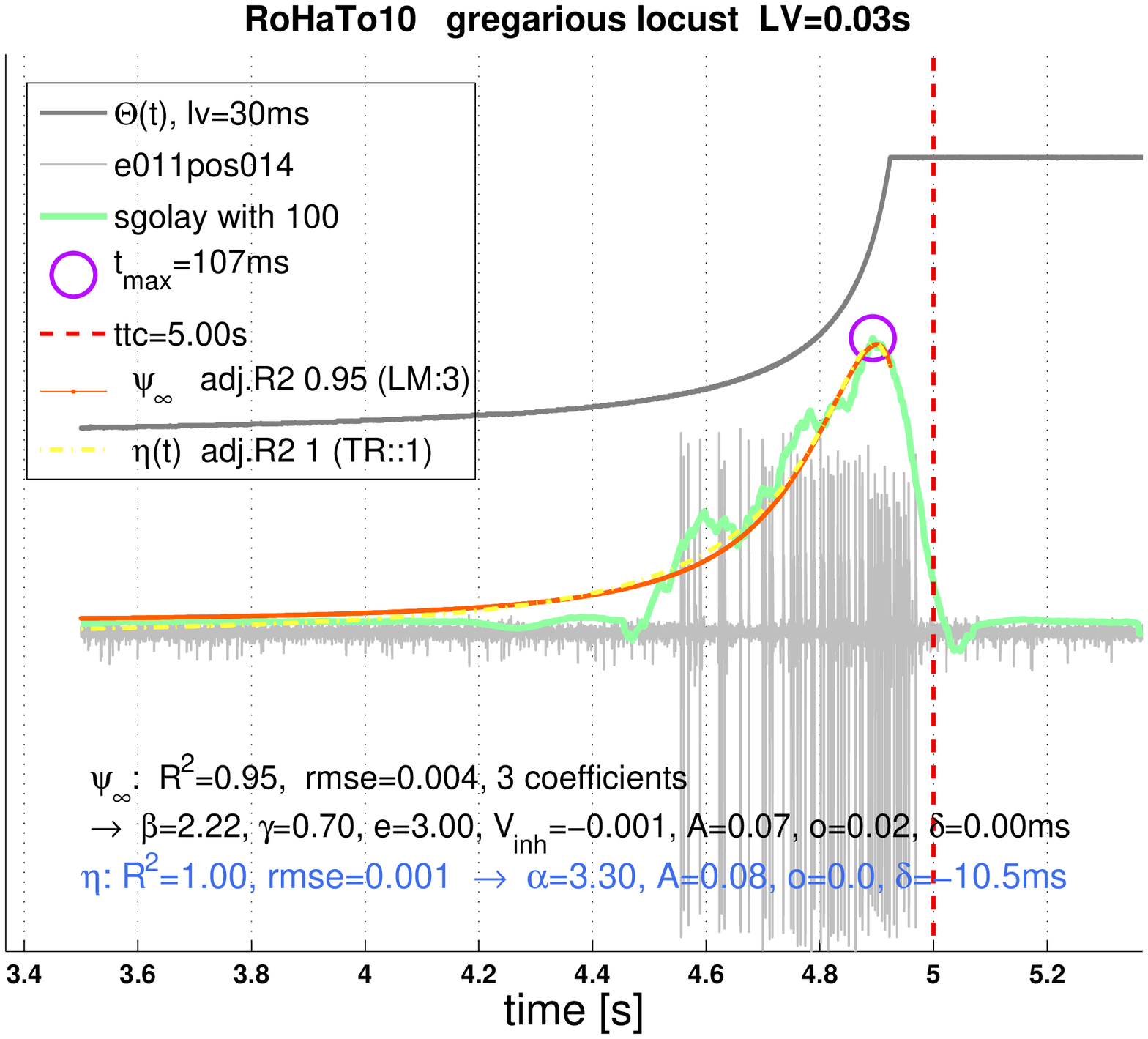}}~
  \subfloat[$\alpha$ versus $\beta$]{\includegraphics[width=0.45\linewidth]{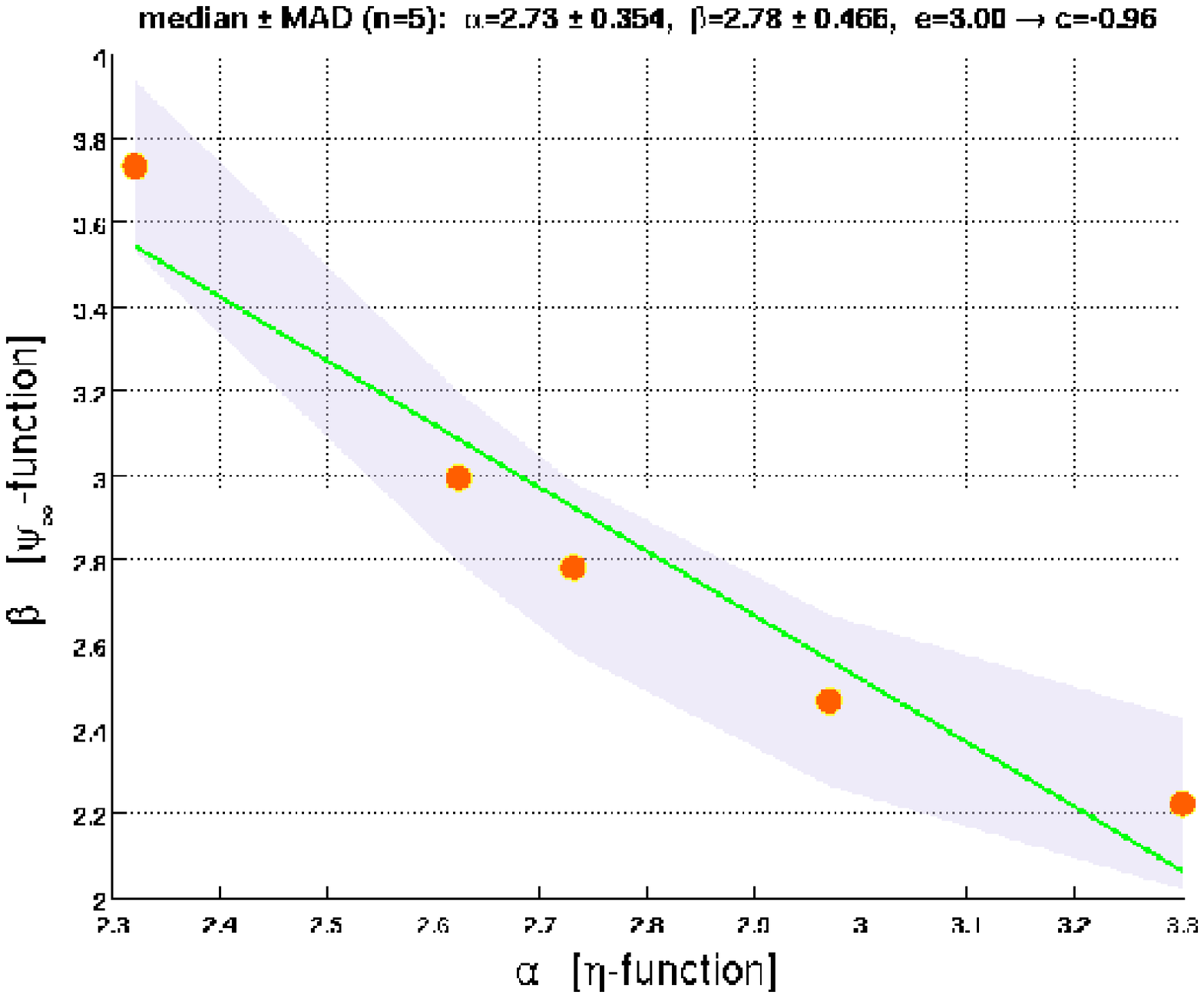}}
\caption{\label{Steve}
\textit{(a)} DCMD activity in response to a black square ($l/|v|=30ms$, legend label ``e011pos14'', ref. \cite{SteveEtAl2010})
approaching to the eye center of a gregarious locust (final visual angle $50^o$).  Data show the first stimulation so habituation
is minimal. The spike trace (sampled at $10^4Hz$) was full wave rectified, lowpass filtered, and sub-sampled to $1ms$ resolution.
Firing rate was estimated with Savitzky-Golay filtering (``sgolay'').  The fits of the $\eta$-function ({\small $A\eta(t+\delta)+o$;
4 coefficients}) and $\psi_{\infty}$-function ({\small $A\psi_{\infty}(t)$ with fixed $e, o, \delta, V_{inh}$; 3 coefficients})
provide both excellent fits to firing rate.
\textit{(b)} Fitting coefficient $\alpha$ ($\rightarrow\eta$-function) inversely correlates with $\beta$ ($\rightarrow\psi_{\infty}$)
when fitting firing rates of another $5$ trials as just described.  Similar correlation values would be obtained if $e$ is fixed
at values $e=2.5,4,5 \rightsquigarrow c=-0.95, -0.96, -0.91$.  If $o$ was determined by the
fitting algorithm, then $c=-0.70$.  No clear correlations with $\alpha$ were obtained for $\gamma$.
}
\end{figure}
%
\section{Freeze! Experimental data versus steady state of ``psi''}
%
In the previous section, experimentally plausible values for $\alpha$ were obtained if $\psi$ is close to equilibrium
at each instant of time during stimulation.  In this section we will thus introduce a steady-state version of $\psi$,
\begin{equation}\label{SteadyStatePsi}
  \psi_{\infty}(t)\equiv\frac{\dot{\Theta}(t) + V_{inh} \left[ \gamma\Theta(t) \right]^e }{\beta + \dot{\Theta}(t) + \left[ \gamma\Theta(t) \right]^e }
\end{equation}
(Here we use continuous versions of angular size and rate of expansion).
The $\psi_{\infty}$-function makes life easier when it comes to fitting experimental data.  However, it has its
limitations, because we brushed the whole dynamic of $\psi$ under the carpet.
Figure \ref{PsiPars} illustrates how the linear relationship (=``linearity'') between $t_{max}\equiv t_c-\hat{t}$ and
$l/|v|$ is influenced by changes in parameter values.  Changing any of the values of $e$, $\beta$, $\gamma$ predominantly causes
variation in line slopes.  The smallest slope changes are obtained by varying $V_{inh}$ (data not shown; we checked
$V_{inh}=0, -0.001,-0.01, -0.1$).  For $V_{inh}\lessapprox -0.01$, linearity is getting slightly compromised, as slope
increases with $l/|v|$ ({\small e.g. $V_{inh}=-1 \rightsquigarrow \alpha\in[4.2,4.7]$}).\\
In order to get a notion about how well the shape of $\psi_{\infty}(t)$ matches $\eta(t)$, we computed time-averaged
difference measures between normalized versions of both functions (details: figure \ref{PsiPars} \& \ref{DiffPsiEta}).
Bigger values of $\beta$ match $\eta$ better at smaller, but worse at bigger values of $l/|v|$ (figure \ref{DiffPsiEta}a).
Smaller $\beta$ cause less variation across $l/|v|$.  As to variation of $e$, overall, curve shapes seem to be best
aligned with $e=3$ to $e=4$ (figure \ref{DiffPsiEta}b).  Furthermore, better matches between $\psi_{\infty}(t)$ and
$\eta(t)$ correspond to bigger values of $\gamma$  (figure \ref{DiffPsiEta}c).
And finally, $V_{inh}$ marches again to a different tune (data not shown). $V_{inh}=-0.1$ leads to the best agreement
($\approx 0.04$  across $l/|v|$) of all $V_{inh}$, quite different from the other considered values.  For the rest,
$\psi_{\infty}(t)$ and $\eta(t)$ align the same (all have maximum $0.094$), despite of covering
different orders of magnitude with $V_{inh}=0, -0.001,-0.01$.\\
\textbf{Decelerating approach}.  Hatsopoulos \emph{et al.} \cite{HatsopoulosEtAl95} recorded DCMD activity
in response to an approaching object which projected image edges on the retina moving at constant velocity:
$\dot{\Theta}=const.$ implies $\Theta(t)=\Theta_0+\dot{\Theta}t$.
This ``linear approach'' is perceived as if the object is getting increasingly slower.  But what
appears a relatively unnatural movement pattern serves as a test for the functions $\eta$ \& $\psi_{\infty}$.
Figure \ref{ConstAngularVelocity} illustrates that $\psi_{\infty}$ passes the test, and consistently predicts
that activity sharply rises in the initial approach phase, and subsequently declines ($\eta$ passed this test
already in the year 1995).\\
\textbf{Spike traces.} We re-sampled about $30$ curves obtained from LGMD recordings from a variety of publications,
and fitted $\eta$ \& $\psi_{\infty}$-functions.  We cannot show the results here, but in terms of goodness of fit measures,
both functions are in the same ballbark.  Rather, figure \ref{Steve}a shows a representative example \cite{SteveEtAl2010}.
When $\alpha$ and $\beta$ are plotted against each other for five trials, we see a strong inverse correlation
(figure \ref{Steve}b).
Although five data points are by no means a firm statistical sample, the strong correlation could indicate that
$\beta$ and $\alpha$ play similar roles in both functions.  Biophysically, $\beta$ is the leakage conductance,
which determines the (passive) membrane time constant $\tau_m \propto 1/\gleak$ of the neuron.  Voltage drops
within $\tau_m$ to its $1/e$ part.  Bigger values of $\beta$ mean shorter $\tau_m$ (i.e., ``faster neurons'').
Getting back to $\eta$, this would suggest $\alpha\propto\tau_m$, such that higher (absolute) values for $\alpha$
would possibly indicate a slower dynamic of the underlying processes.
%
\section{Discussion (``The Good, the Bad, and the Ugly'')}
%
Up to now, mainly two classes of LGMD models existed: The phenomenological $\eta$-function on the one hand,
and computational models with neuronal layers presynaptic to the LGMD on the other (e.g. \cite{RindBramwell96,MatsAngel03gc};
real-world video sequences \& robotics: e.g. \cite{BlaRinVer99,MatsEliAngel04,YueEtAl2006,BermudezBernardetVerschure10}).
Computational models predict that LGMD response features originate from excitatory and inhibitory interactions
in -- and between -- presynaptic neuronal layers.  Put differently, non-linear operations are generated in the
presynaptic network, and can be a function of many (model) parameters (e.g. synaptic weights, time constants, etc.).
In contrast, the $\eta$-function assigns concrete nonlinear operations to the LGMD \cite{GabbianiEtAl2004}.
The $\eta$-function is accessible to mathematical analysis, whereas computational models have to be probed
with videos or artificial stimulus sequences.  The $\eta$-function is vague about biophysical parameters, whereas (good) computational
models need to be precise at each (model) parameter value.  The $\eta$-function establishes a clear link between
physical stimulus attributes and LGMD activity:  It postulates what is to be computed from the optical variables (OVs).
But in computational models, such a clear understanding of LGMD inputs cannot always be expected:  Presynaptic
processing may strongly transform OVs.\\
The $\psi$ function thus represents an intermediate model class: It takes OVs as input, and connects them with
biophysical parameters of the LGMD.  For the neurophysiologist, the situation could hardly be any better.
Psi implements the multiplicative operation of the $\eta$-function by shunting inhibition (equation \ref{membrane}:
$V_{exc}\approx V_{rest}$ and $V_{inh}\approx V_{rest}$).  The $\eta$-function fits $\psi$ very well according
to our dynamical simulations (figure \ref{VoltageTrace}), and satisfactory by the approximate criterion of
figure \ref{DiffPsiEta}.\\
We can conclude that $\psi$ implements the $\eta$-function in biophysically plausible way.  However, $\psi$ does
neither explicitly specify $\eta$'s multiplicative operation, nor its exponential function $\exp(\cdot)$.
Instead we have an interaction between shunting inhibition and a power law $(\cdot)^e$, with $e\approx 3$.
So what about power laws in neurons?\\
Because of $e>1$, we have an expansive nonlinearity.  Expansive power-law nonlinearities are well established in
phenomenological models of simple cells of the primate visual cortex \cite{AlbrechtHamilton1982,Heeger1993}.
Such models approximate a simple cell's instantaneous firing rate $r$ from linear filtering of a stimulus
(say $Y$) by $r \propto ([Y]^+)^e$, where $[\cdot]^+$ sets all negative values to zero and lets all positive pass.
Although experimental evidence favors linear thresholding operations like $r \propto [Y-Y_{thres}]^+$, neuronal
responses can behave according to power law functions if $Y$ includes stimulus-independent noise \cite{MillerTroyer2002}.
Given this evidence, the power-law function of the inhibitory input into $\psi$ could possibly be interpreted as a
phenomenological description of presynaptic processes \cite{BermudezBernardetVerschure10}.


\subsubsection*{Acknowledgments}
MSK likes to thank Stephen M. Rogers for kindly providing the recording data for compiling figure \ref{Steve}.
MSK furthermore acknowledges support from the Spanish Government, by the \emph{Ramon and Cajal} program
and the research grant \emph{DPI2010-21513}.
{\small
\providecommand{\bysame}{\leavevmode\hbox to3em{\hrulefill}\thinspace}
\providecommand{\MR}{\relax\ifhmode\unskip\space\fi MR }
\providecommand{\MRhref}[2]{%
  \href{http://www.ams.org/mathscinet-getitem?mr=#1}{#2}
}
\providecommand{\href}[2]{#2}

}
\end{document}